\documentclass[aps,preprint,amsmath,showpacs,showkeys]{revtex4}

\usepackage{bm}
\usepackage{graphicx}
\usepackage{epsfig}

\begin{document}

\title{Some results for the wave function at the origin for
$\bm{S}$-wave levels}

\author{V.~Gupta}

\email[]{virendra@mda.cinvestav.mx}

\affiliation{
Departamento de F\'{\i}sica Aplicada.\\
Centro de Investigaci\'on y de Estudios Avanzados del IPN.\\
Unidad M\'erida.\\
A.P. 73, Cordemex.\\
M\'erida, Yucat\'an, 97310. MEXICO.
}

\author{G.~S\'anchez-Col\'on}

\email[]{gsanchez@mda.cinvestav.mx}

\affiliation{
Departamento de F\'{\i}sica Aplicada.\\
Centro de Investigaci\'on y de Estudios Avanzados del IPN.\\
Unidad M\'erida.\\
A.P. 73, Cordemex.\\
M\'erida, Yucat\'an, 97310. MEXICO.
}

\date{\today}

\begin{abstract}

Starting with the $S$-wave radial equation for an attractive
central potential $V(r)$, we give results for the $n$ (principal
quantum number) and the $\mu$ (reduced mass) dependence of
$R_{n0}(0)$, the $S$-wave radial wavefunction at the origin, for
potentials with definite curvature.

\end{abstract}


\maketitle

\section{\label{introduction}Introduction}

Discovery of quark-antiquark atoms like charmonium in 1974 led to general
investigations of the Schr\"{o}dinger equation with a central potential $V(r)$
representing the $q\bar{q}$-potential~\cite{quigg}. The motivation was to
obtain results based on general properties of $V(r)$ like its shape, since its
precise form was then (and still is) unknown.

Some results were obtained for the $S$-wave ($\ell=0$) bound state radial wave
function $R_{n0}(r)$, $n$ the principal quantum number~\cite{martin}.
Specifically, it was shown that $R_{20}(0)$ is larger (smaller) than $
R_{10}(0)$ provided $V(r)$ was everywhere convex (concave), $V''(r)>0$
($V''(r)<0)$. The variation of $ R_{10}(0)$ with the reduced mass $\mu$ was
also related to the curvature of the potential. This was directly proved from
the radial equation for $n=1$~\cite{rosner}. Both types of results mentioned
above were proved for large $n$ using the WKB approximation~\cite{VG1}.

In this paper we show that both types of results follow directly from the
$S$-wave radial Schr\"{o}dinger equation for all $n$. For notational simplicity,
define:

\begin{equation}
S_n(0) = [R_{n0}(0)]^2.
\end{equation}

We will prove that for an attractive central potential $V(r)$ and
$n=1,2,3,\ldots$:

\subsection*{Case (a).}

If $V'(r)>0$ and $V''(r)=0$ for all $r$, then:

\begin{equation}
S_n(0) - S_{n+1}(0) =0 \quad\text{and}\quad
\frac{\partial}{\partial\mu}\left[\frac{1}{\mu}S_n(0)\right]=0.
\label{eq2}
\end{equation}

\subsection*{Case (b).}

If $V'(r)>0$ and $V''(r)<0$ for all $r$ and
$V'(\infty)$ is finite, then:

\begin{equation}
S_n(0) - S_{n+1}(0)>0 \quad\text{and}\quad
\frac{\partial}{\partial\mu}\left[\frac{1}{\mu}S_n(0)\right]>0.
\label{eq3}
\end{equation}

\subsection*{Case (c).}

If $V'(r)>0$ and $ V''(r)>0$ for all $r$ and
$V'(0)$ is finite, then:

\begin{equation}
S_n(0) - S_{n+1}(0)<0 \quad\text{and}\quad
\frac{\partial}{\partial\mu}\left[\frac{1}{\mu}S_n(0)\right]<0.
\label{eq4}
\end{equation}

Case (a) corresponds to an attractive linear potential. This case is exactly
solvable. Indeed, there are well known exactly solvable examples for the
concave (Coulomb potential) and convex (harmonic oscillator) cases which
satisfy  the above inequalities. Explicit solutions of convex power law
potentials $r^k$ with $k>1$ and the concave $\log(r)$ potential satisfy the
above inequalities~\cite{HJW}. With all this evidence at hand we believe that
the above inequalities are really theorems. In the next section we establish the
notation and preliminaries, in Sec.~\ref{nonzerol} we present a result for
non-zero $\ell$, followed by our arguments for the $S$-wave results in
Sec.~\ref{zerol}. A confirmation of the results on the $\mu$ dependence of
$R_{n0}(0)$ via a dimensional analysis is presented in Sec.~\ref{dimensional}.
The concluding section contains some discussion.

\section{\label{notation}Notation and preliminaries}

The radial equation, for a two-body system with reduced mass $\mu$ in an
attractive central potential $V(r)$ for $u_{n\ell}(r) =r R_{n\ell}(r)$ is:

\begin{equation}
-C(\mu) u''_{n\ell}(r)+[W_\ell(r)-E_n]u_{n\ell}(r)=0,
\end{equation}

\noindent
where

\begin{equation}
C(\mu) = \frac{\hbar^2}{2\mu}
\end{equation}

\noindent
and

\begin{equation}
W_\ell(r)=V(r) + C(\mu)\,\frac{\ell(\ell+1)}{r^2} .
\end{equation}

The radial wavefunction $R_{n\ell}(r)$ for energy $E_n$ is real so its modulus
square is the same as its square. Consequently, the inequalities in the
introduction are usually stated for the modulus square. The energy of the bound
state increases with the principal quantum number $n$, thus $E_1<E_2
<E_3\ldots$. The potential obeys the standard restrictions, namely, $\lim_{r\to
0}[r^2 V(r)]=0$. Also, recall that $R_{n\ell}(r)$ behaves as $r^{\ell}$ as $r$
tends to zero. For an attractive force, the asymptotic behaviour ($r\to\infty$)
of the radial wavefunction $u_{n\ell}(r)$ will be like $\exp(-ar)$, $a>0$.

Multiply the radial equation by $u'_{n\ell}$ and integrate from zero to
infinity. The term with $E_n$ gives zero. One integration by parts gives:

\begin{equation}
C(\mu)\,[u'{_{n\ell}(0)}]^2\delta_{\ell\,0} =
{\int_0^\infty}W'_{\ell}(r)u^2_{n\ell}(r)dr.
\label{eq8}
\end{equation}

\noindent
The term $W_\ell(r)u^2_{n\ell}(r)$ from the partial integration does
not contribute. This is obvious for the upper limit $r=\infty$. One has to be
careful at the lower limit $r=0$. However, since $V(r)$ is less singular than
$r^{-2}$ and $u^2_{n\ell}(r)\sim r^{2(\ell+1)}$ as $r\to 0$, the lower limit
also does not contribute. All this is well known. Before specializing to
$S$-wave it is interesting to consider the above equation for non-zero $\ell$.

\section{\label{nonzerol}Result for non-zero $\bm{\ell}$}

In this case, since the left hand side of Eq.~(\ref{eq8}) is zero, the equation
simply says that the expectation value of the effective force $W'_{\ell}(r)$ is
zero. Alternatively, it implies that the expectation value of $V'(r)$ for a
general potential is related to that of $r^{-3}$. Explicitly:

\begin{equation}
\langle V'(r)\rangle_{n\ell} =
2C(\mu)\,\ell(\ell+1)\left \langle\frac{1}{r^3}\right\rangle_{n\ell}.
\end{equation}

\noindent
This general result (probably known personally to many~\cite{VG2}) deserves to
be better known. It is is very  useful. For example, for a Coulomb potential it
immediately gives the correct relation between the expectation values of
$r^{-2}$ and $r^{-3}$.

\section{\label{zerol}$\bm{S}$-wave relations}

For $\ell=0$, Eq.~(\ref{eq8}) reduces to:

\begin{equation}
C(\mu)\,S_n(0) = {\int_0^\infty} V'(r) u^2_{n0}(r)dr = \langle
V'(r)\rangle_{n0}.
\label{eq10}
\end{equation}

\noindent
This is a well-known result and provides the basis for the arguments leading to
the proof of the results given in Sec.~\ref{introduction}.

\subsection*{Case (a). $\bm{V''(r)=0}$ for all $\bm{r}$.}

This is the case of the attractive linear potential $V(r)=\lambda r$. So,
$V'(r)=\lambda$ is a positive constant for all $r$. In this case,
Eq.~(\ref{eq10}) reduces to simply

\begin{equation}
C(\mu)\,S_n(0)=\lambda,
\label{eq11}
\end{equation}

\noindent
since $u_{n0}(r)$ is normalized, that is,

\begin{equation}
\int_0^\infty u^2_{n0}(r)dr= 1.
\end{equation}

\noindent
Thus, in this case $C(\mu)\,S_n(0)$ is a constant (the potential strength),
independent of $\mu$ or $n$ as required. It is well known that the linear
potential is exactly solvable in terms of Airy functions. The above relation
for $C(\mu)\,S_n(0)$ has been noted earlier using the explicit
solutions~\cite{Bose}.

\subsection*{Case (b). $\bm{V''(r)<0}$ and $\bm{V'(r)>0}$ for all
$\bm{r}$, with $\bm{V'(\infty)}$ finite.}

A well-known exactly solvable example of this case is the Coulomb potential.
Perform an integration by parts in Eq.~(\ref{eq10}) to obtain:

\begin{equation}
C(\mu)\,S_n(0)=V'(\infty)-{\int_0^\infty}V''(r)f_{n0}(r)dr,
\label{eq13}
\end{equation}

\noindent
where

\begin{equation}
f_{n0}(r)={\int_0^r}u_{n0}^2(r')dr'.
\label{eq14}
\end{equation}

\noindent
Note that $f_{n0}(\infty)=1$ because the radial wavefunction is normalized. The
term $V'(r)f_{n0}(r)$, from the integration by parts at $r=\infty$ gives
$V'(\infty)$ while that at $r=0$ vanishes. This is because $V(r)$ is less
singular than $r^{-2}$ as $r$ tends to 0 while one expects $f_{n0}(r)\sim r^3$
as $r$ tends to 0 because $u_{n0}^2(r)\sim r^2$. Physically, $f_{n0}(r)$
represents the probability of finding the particle (two-body system) between 0
and $r$.

To prove that $S_n(0)-S_m(0)>0$ for $n<m$, we appeal to the virial theorem. For
$S$-wave levels it states:

\begin{equation}
E_{n0} = \langle U(r)\rangle_{n0} = {\int_0^\infty}U(r)u_{n0}^2(r)dr,
\end{equation}

\noindent
where

\begin{equation}
U(r)=V(r)+\frac{1}{2}rV'(r).
\end{equation}

\noindent
So, if $U(\infty)$ is finite, then an integration by parts gives:

\begin{equation}
-[E_{n0} -E_{m0}]={\int_0^\infty}U'(r)[f_{n0}(r)-f_{m0}(r)]dr >0.
\label{virial}
\end{equation}

\noindent
For $n<m$, the left hand side is always positive, so the integral
is positive. Now, $f_{n0}(r)$ is the probability of finding the
bound particle between 0  and $r$ in the eigenstate with energy
$E_{n0}$. Physically, we expect that for all $r$:

\begin{equation}
f_{n0}(r)\ge f_{m0}(r),\quad {\rm for}\quad n<m.
\end{equation}

\begin{figure}

\centerline{\psfig{file=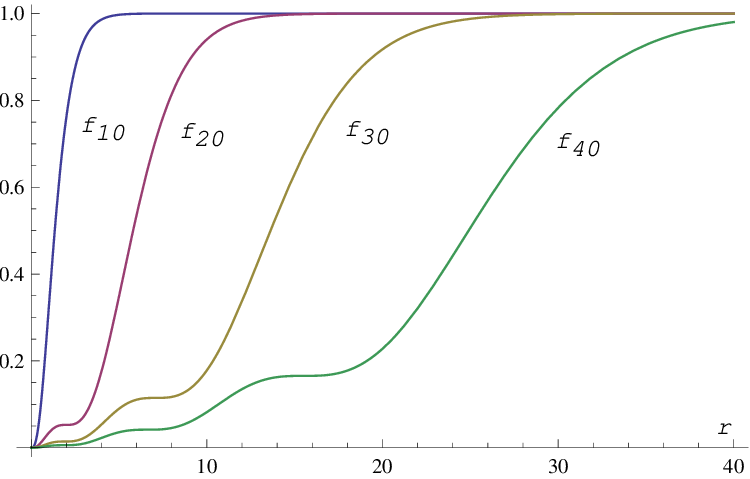,width=6.4in}}

\caption{
\label{Coulomb}
First four lowest $S$-wave levels probabilities, Eq.~(\ref{eq14}), for the
Coulomb potential, showing that $f_{n0}(r)-f_{m0}(r)>0$ for $n<m$ and all $r$.
}

\end{figure}

\noindent
To check this out for the Coulomb potential, $V(r)=-(e/r)$, the
$f_{n0}(r)$ ($n=1,2,3,4$) are plotted in Fig.~\ref{Coulomb}. Thus, it is
clear that if $U'(r) > 0$  for all $r$, then the Virial
theorem~\footnote{Virial theorem  holds for more complicated
potentials, here we are concerned with a limited  class of power
law potentials for which $V\rq(r)$  and $V''(r)$  have the
same sign for all $r$.} is satisfied because $f_{n0}>f_{m0}$
for $n<m$. For example, the power law potentials,
$V(r)=-|\lambda|r^{\alpha}$ with $-2<\alpha<0$ (includes
Coulomb) satisfy that $U'(r) > 0$ and $V''(r)<0$ for all $r$.

Given the above, from Eq.~(\ref{eq13}) it follows that:

\begin{equation}
C(\mu)\,[S_n(0)-S_m(0)]= -{\int_0^\infty}V''(r)[f_{n0}(r)-f_{m0}(r)]dr >0,
\end{equation}

\noindent
for $n<m$ since $V''(r)<0$ for all $r$. This gives the first
inequality in Eqs.~(\ref{eq3}).

For the variation with respect to the reduced mass, we note that
with increasing $\mu$, the bounded system will shrink in size.  So,
physically one expects that $f_{n0}(r)$ will increase, that is,
$\partial[f_{n0}(r)]/\partial\mu >0$. Thus, taking the derivative
with respect to $\mu$ of Eq.~(\ref{eq13}), since $V'(\infty)$ is
a constant, gives the second inequality in Eqs.~(\ref{eq3}).

\subsection*{Case (c). $\bm{V''(r)>0}$ and $\bm{V'(r)>0}$ for all
$\bm{r}$, with $\bm{V'(0)}$ finite.}
 
In this case we perform a slightly different integration by parts
in Eq.~(\ref{eq10}) to obtain:

\begin{equation}
C(\mu)\,S_n(0)= V'(0) + {\int_0^\infty}V''(r) g_{n0}(r)dr,
\label{eq20}
\end{equation}

\noindent
where

\begin{equation}
g_{n0}(r)={\int_r^\infty}u_{n0}^2(r)dr=1-f_{n0}(r).
\end{equation}

\noindent
The term $ V'(r) g_{n0}(r)$, from the integration by parts, gives
$ V'(0)$ for $ r=0$. For the upper limit $r\to\infty$ it vanishes
because $u_{n0}(r)$ represents a bound state.

From the above two equations we obtain:

\begin{eqnarray}
C(\mu)\,[S_n(0)-S_m(0)]&=&{\int_0^\infty}V''(r)[g_{n0}(r)-g_{m0}(r)]dr
\nonumber\\
&=& -{\int_0^\infty}V''(r)[f_{n0}(r)-f_{m0}(r)]dr.
\label{eq22}
\end{eqnarray}

\begin{figure}

\centerline{\psfig{file=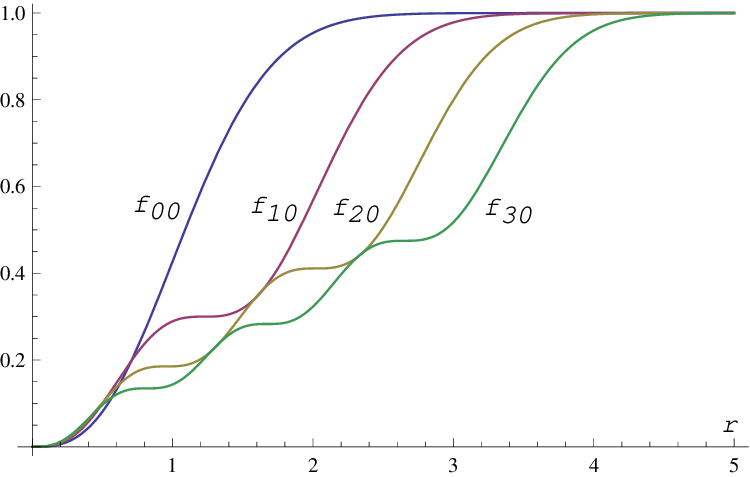,width=6.4in}}

\caption{
\label{HarmOsc}
First four lowest $S$-wave levels probabilities, Eq.~(\ref{eq14}), for the
Harmonic Oscillator potential, showing that in this case
$f_{n0}(r)-f_{m0}(r)\ge0$ for $n<m$ and all $r$ except for a small region near
the origin.
}

\end{figure}

\begin{figure}

\centerline{\psfig{file=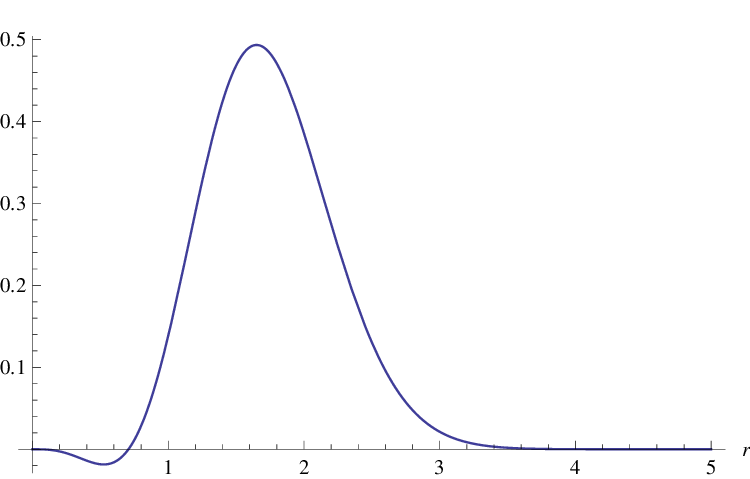,width=6.4in}}

\caption{
\label{f00mf10}
The difference $f_{00}(r)-f_{10}(r)$ for the Harmonic Oscillator potential,
showing that in this case $f_{n0}(r)-f_{m0}(r)$ is slightly negative for
small $r$ in a small region near the origin and after that it is positive
for all $r$ for $n<m$.
}

\end{figure}

\begin{figure}

\centerline{\psfig{file=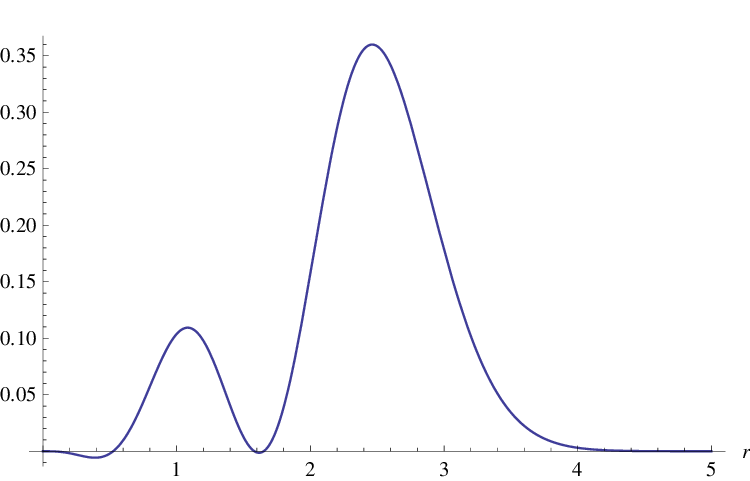,width=6.4in}}

\caption{
\label{f10mf20}
The difference $f_{10}(r)-f_{20}(r)$ for the Harmonic Oscillator potential,
showing that in this case $f_{n0}(r)-f_{m0}(r)$ is slightly negative for
small $r$ in a small region near the origin and after that it is positive
for all $r$ for $n<m$. The negative region is smaller as $n$ increases.
}

\end{figure}

\begin{figure}

\centerline{\psfig{file=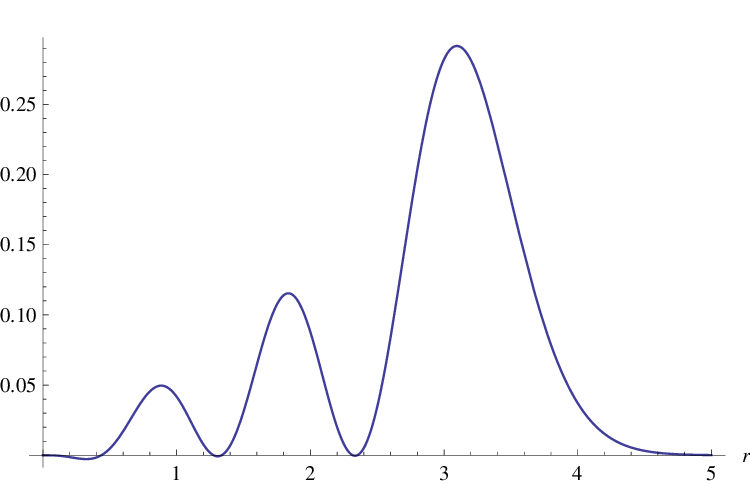,width=6.4in}}

\caption{
\label{f20mf30}
The difference $f_{20}(r)-f_{30}(r)$ for the Harmonic Oscillator potential,
showing that in this case $f_{n0}(r)-f_{m0}(r)$ is slightly negative for
small $r$ in a small region near the origin and after that it is positive
for all $r$ for $n<m$. The negative region is smaller as $n$ increases.
}

\end{figure}

\noindent
Power law potentials $V(r)=|\lambda|r^{\alpha}$ with $\alpha>1$ satisfy the
conditions for the above results. In particular, $\alpha=2$ gives the isotropic
harmonic oscillator which is exactly soluble. Figure~\ref{HarmOsc} gives plots
of $f_{n0}(r)$ for $n=0,1,2,3$. The differences $f_{00}(r)-f_{10}(r)$,
$f_{10}(r)-f_{20}(r)$, and $f_{20}(r)-f_{30}(r)$, are plotted in
Figs.~\ref{f00mf10}--\ref{f20mf30}, respectively. Unlike the Coulomb case, in
this case the differences are slightly negative for small $r$ in a small region
near the origin and after that they are positive for all $r$. Even so, the
integrals ${\int_0^\infty}r[f_{n0}(r)-f_{m0}(r)]dr$ (in the Virial theorem,
Eq.~(\ref{virial})) and ${\int_0^\infty}[f_{n0}(r)-f_{m0}(r)]dr$ (in
Eq.~(\ref{eq22})) are positive. Consequently, physically one expects that the
integrals ${\int_0^\infty}r^{\alpha-1}[f_{n0}(r)-f_{m0}(r)]dr$ and
${\int_0^\infty}r^{\alpha-2}[f_{n0}(r)-f_{m0}(r)]dr$ will be positive for power
law potentials $V(r)=|\lambda|r^{\alpha}$ with $\alpha\ge 2$. This is supported
by explicit solutions for convex power law potentials with
$\alpha>1$~\cite{HJW}. Since $V''(r)>0$, this analysis leads to $S_n(0) -
S_m(0)<0$, the first inequality in Eqs.~(\ref{eq4}).

For the variation with respect to the reduced mass $\mu$, we
obtain from Eq.~(\ref{eq20}):

\begin{equation}
\frac{\partial}{\partial\mu}\left[\frac{1}{\mu}S_n(0)\right]<0,
\end{equation}

\noindent
since the variation with $\mu$ of $g_{n0}(r)$ is opposite to that
of $f_{n0}(r)$.  This concludes the proof of the $S$-wave results
given in Sec.~\ref{introduction}.

\section{\label{dimensional}Dimensional analysis confirmation of the variation
of $\bm{\mu^{-1}S_n(0)}$  with reduced mass
$\bm{\mu}$.}
 
Consider the power law potential

\begin{equation}
V(r)=\lambda r^{\alpha},
\end{equation}

\noindent
sign of $\lambda$  is chosen depending on the range of $\alpha$ so that $V'(r)$
is positive and $V(r)$ has bound states. For example, for the Coulomb potential
$\alpha=-1$ and $\lambda<0$.

Typical length scale, $a_0$,  for bound states will depend on
$\lambda$, $\hbar$,  and the reduced mass $\mu$, the parameters
in the Schr\"{o}dinger equation. Dimensional analysis gives

\begin{equation}
a_0\sim\left(\frac{\hbar^2}{|\lambda|\mu}\right)^{\frac{1}{2+\alpha}},
\end{equation}

\noindent
For Coulomb case $\lambda=-e^2$, $\alpha=-1$, so $a_0$ is just
the Bohr radius.

Since the wavefunction is normalized $S_n(0)$ has dimensions of
(length)$^{-3}$ so,  dimensionally,

\begin{equation}
\frac{1}{\mu}S_n(0)\sim\frac{1}{\mu}
\left(\frac{|\lambda|\mu}{\hbar^2}\right)^{\frac{3}{2+\alpha}},
\label{eq26}
\end{equation}

\noindent
this formula gives the required dependence on $\mu$ for the
various power law potentials. We apply it to the cases treated
earlier.

\subsection*{Case (a).}

For the linear potential $\lambda>0$ and
$\alpha=1$, there is no $\mu$ dependence, so

\begin{equation}
\frac{\partial}{\partial\mu}\left[\frac{1}{\mu}S_n(0)\right]
=0,
\end{equation}

\noindent
in agreement with Eqs.~(\ref{eq2}) and (\ref{eq11}).

\subsection*{Case (b).}

For potentials where
$V(r)=-|\lambda|{r}^{\alpha}$ with $-2<\alpha<0$, the power of
$\mu$ in Eq.~(\ref{eq26}) is positive, in accord with
Eq.~(\ref{eq3}).

\subsection*{Case (c).}

For potentials where
$V(r)=|\lambda|{r}^{\alpha}$ with $1<\alpha$, the power of $\mu$
in Eq.~(\ref{eq26}), $(1-\alpha)/(2+\alpha)$, is negative, in
accord with Eq.~(\ref{eq4}).

\section{\label{conclusions}Concluding remarks}

The proofs of the $S$-wave results presented above have appealed to the physical
meaning of the quantities involved and how they are expected to change
physically with the energy of the bound state (or $n$) and the reduced mass
$\mu$. All known soluble examples of attractive potentials with curvature of
the same sign for all $r$ support the results given in the introduction. Such
potentials imply that the bound states lie in a single potential well. This is
important for the physical arguments presented here. A potential with more than
a single well cannot possibly have curvature of the same sign everywhere. There
are lot of solvable potentials for $S$-waves~\cite{GP} without definite
curvature for which the results given in the Introduction may or may not hold.
It would be an interesting challenge to find a counter example to the
inequalities presented in this work.

\begin{acknowledgments}

It is a pleasure to thank Antonio Bouzas and Andr\'{e}s
G.~Saravia for discussions and help. The authors would like to
thank CONACyT (M\'exico) for partial support.

\end{acknowledgments}

\clearpage

\end{document}